\documentclass[12pt]{article}
\usepackage{graphicx,latexsym}

  \def\pd{\partial} \def\pp{\prime}  \def\b{\beta} \def\dl{\delta} \def\s{\sigma}  \def\vphi{\varphi} \def\eps{\epsilon} 
 \def\lam{\lambda} \def\Lam{\Lambda} \def\gm{\gamma} 
 \def\Om{\Omega}  \def\sq{\sqrt} \def\e{\hbox{\large \it e}}
 \def\fr{\frac}  

\def\P{{\rm P}} \def\QG{{\rm QG}} \def\pl{{\rm pl}} \def\M{{\rm M}} \def\T{{\rm T}}

\def\hphi{{\hat \phi}}

 \def\bg{{\bar g}} \def\bDelta{{\bar \Delta}}

  \def\bx{{\bf x}}  \def\bk{{\bf k}} 
 \def\bq{{\bf q}}    
\def\D{{\bf D}}

\def\lap3{~| \!\!\! \partial^2} \def\dlap3{~| \!\!\! \partial^4}

\def\dH{{\dot H}} \def\ddH{{\ddot H}} \def\dddH{\stackrel{...}{H}}

\begin{document}

\begin{titlepage}


\vspace{3mm}

\begin{center}
{\Large {\bf From CFT Spectra to CMB Multipoles in Quantum Gravity Cosmology}}
\end{center}

\vspace{3mm}

\begin{center}
{\sc Ken-ji Hamada$^1$, Shinichi Horata$^2$ and Tetsuyuki Yukawa$^3$}
\end{center}

\begin{center}
{}$^1${\it Institute of Particle and Nuclear Studies, KEK, Tsukuba 305-0801, Japan} \\
{}$^1${\it Department of Particle and Nuclear Physics, The Graduate University for Advanced Studies (Sokendai), Tsukuba 305-0801, Japan} \\
{}$^{2,3}${\it Hayama Center for Advanced Studies, The Graduate University for Advanced Studies (Sokendai), Hayama 240-0193, Japan} \\
\end{center}

\begin{abstract}
We study the inflation process of universe based on the renormalizable quantum gravity formulated as a conformal field theory (CFT). We show that the power-law CFT spectrum approaches to that of the Harrison-Zel'dovich-Peebles type as the amplitude of gravitational potential gradually reduces during the inflation. The non-Gaussanity parameter is preserved within order of unity due to the diffeomorphism invariance. Sharp fall-off of the angular power spectrum of cosmic microwave background (CMB) at large scale is understood as a consequence of the existence of dynamical scale of the quantum gravity $\Lam_\QG (\simeq 10^{17}{\rm GeV})$. The angular power spectra are computed and compared with the WMAP5 and ACBAR data with a quality of $\chi^2/{\rm dof}\simeq 1.1$.
\end{abstract}

\end{titlepage}

\section{Introduction}
\setcounter{equation}{0}
\noindent

Recent observations of anisotropies in the cosmic microwave background (CMB) by various groups such as the cosmic background explore (COBE) \cite{cobe}, the Wilkinson microwave anisotropy probe (WMAP) \cite{wmap,wmap5}, and the arcminute cosmology bolometer array receiver (ACBAR) \cite{acbar} have provided a refined picture of the history of universe after the big bang. Cosmological parameters are determined with high accuracy based on the cosmological perturbation theory \cite{bardeen,kodama,hs,ll}, assuming only the primordial spectrum close to that of the Harrison-Zel'dovich-Peebles \cite{harrison,zeldovich,py}. We believe that one of the important problems remained in the study of inflation \cite{guth,sato,linde,as,starobinsky,mc,starobinsky2,vilenkin,hhr} is to clarify dynamics producing such a scale-invariant spectrum from the fundamental theory rather than introducing  an artificial field by hands just for the phenomenological purpose.

As the fundamental theory, we will employ the renormalizable quantum gravity formulated based on the conformal field theory (CFT) in four dimensions \cite{hamada02,hamada09,nova,hamada08}. It predicts that quantum fluctuations of the conformal mode in gravitational fields become so large at very high energies beyond the Planck scale, and a conformally invariant space-time is realized as a consequence of background metric independence. It then produces a power-low spectrum and a non-Gaussian fluctuation distribution for the theoretical generation of CMB spectrum.

Evolution of the early universe can be regarded as a violating process of conformal invariance \cite{hy,nova,hhy,hhsy}. The conformal symmetry starts to be broken at the Planck scale, and the space-time dynamics shifts to the inflationary epoch with the expansion time constant about the Planck mass $m_\pl(=1/\sq{G})$. The conformal invariance is completely broken at the dynamical scale of quantum gravity $\Lam_\QG$, which is expected to be $10^{17}$GeV. At this energy scale, the inflation terminates, and the universe turns to the classical Friedmann universe.

The purpose of this paper is to clarify how the Harrison-Zel'dovich-Peebles spectrum is prepared for the initial condition of the cosmological perturbation equation for computing the CMB angular power spectra. The time evolution of gravitational fluctuations in the inflationary background has been studied within the linear approximation \cite{hhy}. It was shown that during the inflation the amplitude of scalar fluctuation decreases to the size which solves the flatness problem. Combining this result with smallness of the non-Gaussianity, we will show that the `almost' Harrison-Zel'dovich-Peebles spectrum (a constant spectrum with rapid fall-off at small momenta) emerges after the inflation.

The tensor mode which measures a degree of deviation from the conformal invariance is expected to be small initially because of the asymptotically free behavior of this mode, while its amplitude is preserved during the inflation. Thus, the tensor mode also gives a significant contribution at the later stage in the primordial spectra for the computation of the CMB multipole distribution.

The correlation length of quantum gravity is given by the order of $\xi_\Lam=1/\Lam_\QG$, and it brings the absence of correlations of two points separated larger than $\xi_\Lam$ initially prepared before the universe starts inflation. This explains the sharp fall off of the angular power spectra at low multipoles \cite{hy}.

\section{Quantum Gravity Cosmology}
\noindent

The renormalizable quantum gravity formulated as a perturbed theory from CFT is defined by the dimensionless action \cite{hamada02, hamada09, nova, hamada08},
\begin{equation}
   I = \int d^4 x \hbox{$\sq{-g}$} \left\{
      -\fr{1}{t^2} C_{\mu\nu\lam\s}^2 -b G_4 
       + \fr{1}{\hbar} \left( \fr{1}{16\pi G}R -\Lam + {\cal L}_\M \right) \right\}, 
       \label{action}
\end{equation}
where $C_{\mu\nu\lam\s}$ is the Weyl tensor, $G_4$ is the Euler density, and $t$ is a dimensionless coupling constant which measures a degree of deviation from CFT. The cosmological constant is denoted by $\Lam$, whose effect can be neglected in the early universe. ${\cal L}_\M$ represents the Lagrangean for conformally invariant matter fields, and $\hbar$ is the Planck constant which is taken to be unity.

The metric field is decomposed to the conformal mode $\phi$ and the traceless tensor mode $h_{\mu\nu}$ as
\begin{equation}
     g_{\mu\nu}=e^{2\phi}\bg_{\mu\nu}, \qquad
     \bg_{\mu\nu}=\eta_{\mu\nu} + h_{\mu\nu},
       \label{metric decomposition}
\end{equation}
with $tr(h)=0$. Signature of the flat background metric $\eta_{\mu\nu}$ is $(-1,1,1,1)$, which defines the conformal time and the comoving frame with the coordinates $x^\mu=(\eta,x^i)$ and $\pd_\mu =(\pd_\eta, \pd_i)$.

The traceless tensor mode, which is governed by the Weyl action, is handled perturbatively in terms of the coupling $t$. The renormalized coupling constant $t_r$ indicates the asymptotic freedom with the beta function $\b_t=-\b_0 t_r^3~(\b_0 >0)$ \cite{ft,hamada02}. It justifies the perturbative treatment of this mode, and also implies the existence of a dynamical energy scale $\Lam_\QG$. The running coupling constant is then written as $1/\bar{t}^2_r(p)=\b_0 \log (p^2/\Lam_\QG^2)$ for a physical momentum $p$.

A recent significant progress in quantization techniques is that the conformal mode has been managed non-perturbatively, as in the case of two-dimensional quantum gravity \cite{polyakov,kpz,dk}. When we rewrite the diffeomorphism invariant measure in terms of the practical measures defined on the flat background metric $\eta_{\mu\nu}$, the partition function is expressed as $Z=\int [d\phi dh \cdots]_\eta \exp ( iS+iI )$. The induced action $S$ is the Jacobian needed to recover the diffeomorphism invariance. At the lowest order of the coupling $t_r$, it gives the kinetic term of the conformal mode, called the Riegert action \cite{riegert,am,amm92,hamada99},
\begin{equation}
   S = -\fr{b_1}{8\pi^2} \int d^4 x \phi \bDelta_4 \phi, 
           \label{Wess-Zumino action}
\end{equation}
where $\bDelta_4=(\pd_\lam \pd^\lam)^2+o(h)$ is a conformally invariant fourth-order differential operator for a scalar field variable. This action is a four-dimensional counter part of the Liouville-Polyakov action in two dimensions. The coefficient has been computed to be $b_1 = ( N_X + 11N_W/2 + 62 N_A )/360 + 769/180$, where $N_X$, $N_W$, and $N_A$ are numbers of scalar fields, Weyl fermions, and gauge fields, respectively, and the last term is a quantum loop correction from gravitational fields. Since $b_1$ for various GUT models is given about $10$, we consider this case in the following.

There are various proporsals in the past how to treat the problem of ghosts. For example, Tomboulis \cite{tomboulis} proposed that ghosts might be removed in the IR region based on the idea of Lee and Wick using the resummed propagator in asymptotically free theories. The idea of asymptotic safety by Weinberg \cite{weinberg} is defined as a cutoff model expanded in derivatives about the ghost-free Einstein theory assuming the existence of a non-trivial UV fixed point where the cutoff is taken to be infinity. Recently, Horava \cite{horava} proposed a higher-derivative gravity model by making ghosts non-dynamical at the cost of diffeomorphism invariance in the UV limit.

Our proposal \cite{hamada08} is that the problem of ghosts should be reconsidered under the light of CFT described by the combined system of the Riegert and the Weyl actions which appears in the UV limit. Since the conformal symmetry realized as a part of diffeomorphism invariance mixes positive-metric and negative-metric modes of the gravitational field, we cannot consider these modes separately and thus the field acts as a whole in physical quantities. At present we do not have a complete proof on the unitarity problem yet, but there is no unphysical behavior at least within  discussions given in this paper.

The quantum gravity cosmology \cite{hy,nova,hhy,hhsy,hms} is now defined by the effective action described by $S+I$ together with higher $t_r$ corrections in which the two mass scales are ordered as $m_\pl \gg \Lam_\QG$. When the universe is at the Planck scale, the coupling constant is negligibly small due to the asymptotic freedom, and the equation of motion for the homogeneous component of conformal mode, $\hphi$, is obtained to be 
\begin{equation}
 b_1 \pd_\eta^4 \hphi - 3\pi m_\pl^2 \e^{2\hphi} ( \pd_\eta^2 \hphi + \pd_\eta \hphi \pd_\eta \hphi ) =0.
      \label{homogeneous equation}
\end{equation}
We introduce the proper time $\tau$ through $d\tau = a d\eta$ with the scale factor $a=e^\hphi$. In terms of $H=\dot{a}/a$, where dot represents the proper time derivative, the stable solution is written as $H=H_\D$ which corresponds to the inflationary universe $a(\tau) \propto e^{H_\D \tau}$, where $H_\D=m_\pl\sq{\pi/b_1}$. The constant $H_\D$ has a order of the Planck scale, and $\tau_\P=1/H_\D$ is called Planck time in the followings.

As the universe expands, the running coupling constant grows and diverges at the dynamical time scale $\tau_\Lam (=1/\Lam_\QG)$. At this point,  the conformal symmetry completely breaks down and quantum coherence of fields disappears resulting the emergence of classical space-time.

\begin{figure}[h]
\begin{center}
\includegraphics[width=8cm,clip]{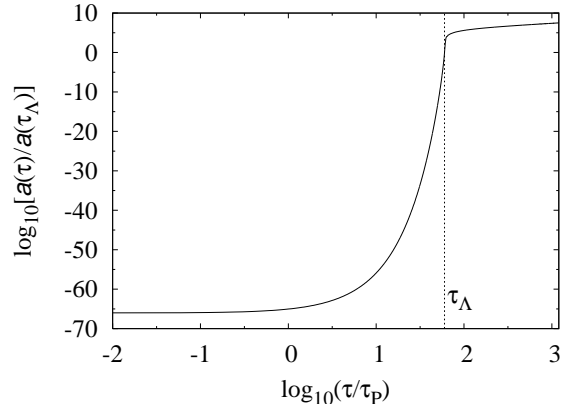}
\end{center}
\caption{\label{evolution of scale factor} {\small Evolution of the scale factor $a(\tau)$. The space-time starts growing rapidly at the Planck time $\tau_\P$. The inflation ends at the dynamical time scale $\tau_\Lam ~(=60 \tau_\P)$. After $\tau$ bigger than $\tau_\Lam$, the scale factor follows the equation of motion from the low energy effective theory of gravity given by an expansion in derivatives of the metric field in which the lowest term is the Einstein action (see \cite{hhy} in detail).}}
\end{figure}

The evolution equation incorporating such a space-time dynamics \cite{hhy} is effectively obtained by including quantum corrections to the coefficient $b_1$ in equation (\ref{homogeneous equation}) by $b_1(1-a_1 t_r^2+\cdots)=b_1B_0(t_r)$ \cite{nova,hamada09}. The higher order corrections are taken into account in the dynamical factor $B_0$ by assuming a resummention form
\begin{equation}
    B_0(t_r) = \fr{1}{(1+ \fr{a_1}{\kappa}t_r^2)^\kappa} .
    \label{dynamical factor}
\end{equation}
By regarding the energy scale to be the inverse of time in the expanding universe, the time-dependence of $B_0$ is defined  by replacing $t^2_r$ to the running coupling constant $\bar{t}^2_r(\tau) = 1/\b_0 \log(1/\tau^2 \Lam_\QG^2)$, which is a solution of the renormalization group equation $-\tau d\bar{t}_r/d\tau=\b_t(\bar{t}_r)$. The equation of motion is then written in terms of the Hubble variable as
\begin{equation}
     B_0(\tau) \left( \dddH +7H\ddH +4\dH^2 +18H^2\dH +6H^4 \right)
     -3 H_\D^2 \left( \dH +2H^2 \right) =0 .
     \label{homogeneous evolution equation}
\end{equation} 
It shows that when the running coupling diverges at $\tau_\Lam$, the dynamical factor $B_0$ vanishes, and the transition from the conformal gravity to the Einstein gravity occurs, as shown in fig.\ref{evolution of scale factor}.\footnote{ 
The parameters, $\b_0$, $a_1$ and $\kappa$, are determined phenomenologically because they depend on strong-coupling dynamics of the traceless tensor mode. In this paper, they are chosen to be $\b_0/b_1 =0.06$, $a_1/b_1=0.01$ and $\kappa=0.5$ as used in \cite{hhy}. 
} 

Similarly, we obtain the equation of energy conservation from the time-time component of the stress tensor as
\begin{equation}
     B_0(\tau) \left( 2H\ddH -\dH^2 +6H^2\dH +3 H^4 \right) -3 H_\D^2 H^2 
        + 8\pi^2\rho/b_1 =0,
\end{equation}
where $\rho$ is the matter density. This equation shows that the matter density vanishes initially because of $H \simeq H_\D$ and $B_0 \simeq 1$ at $\tau_\P$, while it increases sharply as $B_0$ vanishes at $\tau_\Lam$. The energy which creates matter fields $\rho$ originates from a degree of freedom in the conformal mode.

\begin{figure}[h]
\begin{center}
\includegraphics[scale=0.7]{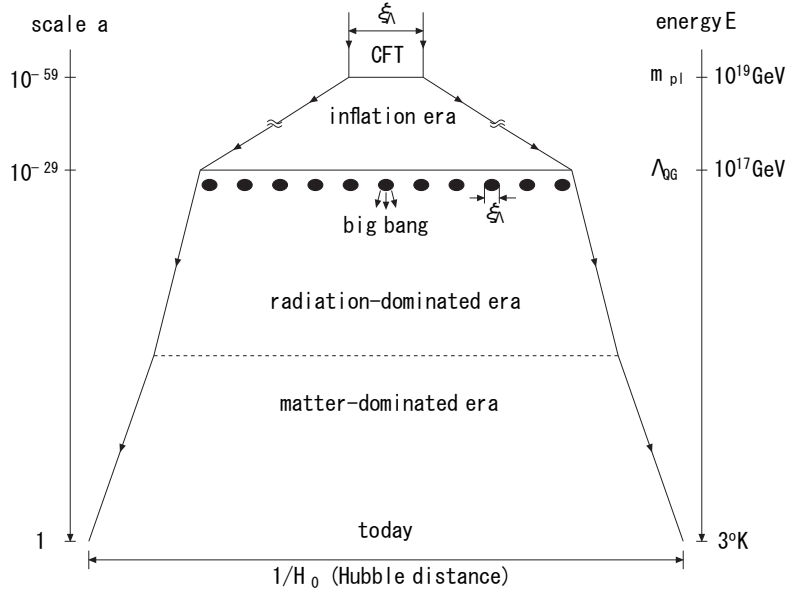}
\end{center}
\caption{\label{evolution scenario} {\small Evolution scenario of the universe based on the quantum gravity cosmology. The initial fluctuation prior to inflation with the correlation length $\xi_\Lam =1/\Lam_\QG ~(\gg L_\P)$ is expanded up to the Hubble distance $1/H_0 (\simeq 5000{\rm Mpc})$ today. There is no correlation between two points with a distance larger than $\xi_\Lam$. Therefore, two points separated beyond $\xi_\Lam$ multiplied by the scale factor $10^{59}$ today lack correlation, and it is our resolution on the sharp fall-off of CMB angular power spectra at large angles.} }
\end{figure}

The number of e-foldings from the Planck time $\tau_\P$ to the dynamical time $\tau_\Lam$ is approximately given by the ratio of two mass scales: ${\cal N}_e =\log[a(\tau_\Lam)/a(\tau_\P)] \sim H_\D/\Lam_\QG$. We here choose $H_\D/\Lam_\QG=60$ which is apropriate to solve the flatness problem, and then the dynamical scale is given by $\Lam_\QG \simeq 10^{17}$GeV. The evolution scenario of the universe based on the quantum gravity cosmology is depicted in fig.\ref{evolution scenario}.

Amplitude of the relative scalar curvature fluctuation, $\dl_R = \dl R/R$, in the inflationary background is then estimated as follows. Since the curvature has two derivatives, the curvature fluctuation would be order of square of the energy scale. Hence, $\dl_R$ is about $\Lam^2_\QG/12H_\D^2 \sim 1/12{\cal N}_e^2$ at the transition point, where the denominator is the curvature of inflationary background. It gives the magnitude about the order of $10^{-5}$ consistent to the observed value by the COBE and WMAP.

In this way, we revive Starobinsky's idea of inflation \cite{starobinsky,mc,starobinsky2,vilenkin,hhr} on the foundation of modern quantum field theory.

\section{Scale Invariant Spectra}
\noindent

Let us describe evolution of the conformal mode starting from the period before inflation to the transition point. Our main concern is to show how the initial power law spectrum propagates to the effective spectrum at the transition point, which serves as the initial spectrum for the classical equation of cosmological perturbation theory.

\paragraph{Scalar spectrum}
The initial spectrum is set at a time $\tau_i =1/E_i$ with $E_i \geq H_\D$ before the space-time starts inflation, where the scale factor is considered to stay almost constant (see fig.\ref{evolution of scale factor}). The scalar spectrum at this epoch is then given by the correlation function of the scalar curvature operator governed by the Riegert action (\ref{Wess-Zumino action}). Denoting $\vphi$ to be a fluctuation of the conformal mode, the relative scalar curvature fluctuation is written as 
\begin{equation}
     \dl_R = \fr{\dl R}{12m^2} = \fr{1}{2m^2} e^{2\vphi}\left( - \pd_i \pd^i \vphi - \pd_i \vphi \pd^i \vphi \right)
\end{equation}
with the initial conditions $\pd_\tau \vphi=\pd_\tau^2 \vphi =0$. Here, $m = a(\tau_i)H_\D$ is the Planck mass at $\tau_i$ in the comoving frame, where the scale factor of today is normalized to be unity. The Fourier transform of this operator in the comoving momentum is given by
\begin{equation}
    {\tilde \dl}_R (k) = \fr{k^2}{2m^2} {\tilde \vphi}_{\rm NL}(k),
       \label{definition of vphi_NL}
\end{equation}
where 
\begin{equation}
    {\tilde \vphi}_{\rm NL}(k) =  {\tilde \vphi}(k) 
      + \int \fr{d^3\bq}{(2\pi)^3} {\tilde \vphi} \left( \bk/2-\bq \right) 
                                   {\tilde \vphi} \left( \bk/2 + \bq \right) 
         \left( \fr{3}{4}+\fr{q^2}{k^2} \right) 
      \label{expansion of vphi_NL}
\end{equation}
up to the second order of $\vphi$. This expression shows that the so-called local non-Gaussianity or non-linearity parameter $f_{\rm NL}$ of the scalar spectrum defined by $\vphi_{\rm NL}(\bx) =\vphi(\bx)+f_{\rm NL}\vphi^2(\bx)$ in the real space \cite{ks} is of the order unity, $f_{\rm NL} \simeq 1$.

We begin with considering the linear part of the initial scalar spectrum. Prior to the inflation the four-derivative dynamics of the action (\ref{Wess-Zumino action}) brings a long range correlation much larger than the horizon distance of the inflationary phase given by the Planck length $L_\P=1/H_\D$. The equal time two-point correlation function of the $\vphi$ field is written in terms of the logarithmic function as
\begin{equation}
   \langle \vphi(\tau_i,\bx)\vphi(\tau_i,\bx^\pp) \rangle = -\fr{1}{4b_1} \log \left( m^2 |\bx-\bx^\pp|^2) \right)
       \label{log correlation}
\end{equation}
at $\tau_i$. The Fourier transform of the logarithmic correlation is given by the formula
\begin{eqnarray}
    -\log \left( m^2|\bx|^2 \right) 
    = \int_{k >\epsilon} \fr{d^3 \bk}{(2\pi)^3}  \fr{4\pi^2}{k^3} e^{i\bk \cdot \bx} 
       - \log \left( \fr{m^2}{\eps^2 e^{2\gm-2}} \right),
\end{eqnarray}
where $\eps$ is an infinitesimal cut-off and $\gm$ is the Euler constant. Since the constant term in Fourier space is proportional to $\dl^3(\bk)$, we disregard it. Thus, we obtain the scale-invariant spectrum of Harrison-Zel'dovich-Peebles as the initial spectrum, 
\begin{equation}
    \fr{k^3}{2\pi^2} \langle |{\tilde \vphi}(\tau_i,\bk)|^2 \rangle = \fr{1}{2b_1} ,
      \label{linear spectrum}
\end{equation}
where $\langle {\tilde \vphi}(\bk){\tilde \vphi}(\bk^\pp) \rangle = \langle |{\tilde \vphi}(\bk)|^2 \rangle (2\pi)^3 \dl^3(\bk+\bk^\pp)$. The positivity of the amplitude reflects the right sign of the Riegert action, $b_1 >0$.

\begin{figure}[h]
\begin{center}
\includegraphics[width=10cm,clip]{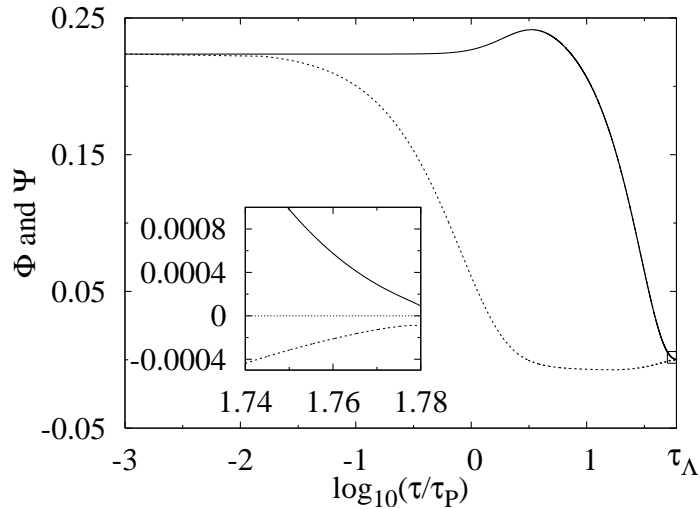}
\end{center}
\caption{\label{evolution of scalar mode} {\small The evolution of the Bardeen potentials, $\Phi$(solid) and $\Psi$(dashed), in the inflationary background within the linear approximation. The initial value satsifying $\Phi=\Psi(=\vphi)$ is taken to be the constant $1/\sq{2b_1}$ with $b_1=10$ and $k=0.01$Mpc$^{-1}$, provided $m=0.0156~(=60\lam)$Mpc$^{-1}$. The other parameters are taken to be those used in \cite{hhy}. The Bardeen potentials reduce their amplitudes, satisfying $\Phi=-\Psi$ at the transition point.}}
\end{figure}

In order to obtain the spectrum at the transition, we have to solve the evolution equation for fluctuations. We first consider linear equations of motion with the initial spectrum (\ref{linear spectrum}) and see the inflationary solution is stable. The evolution equations are given in Appendix.\footnote{ 
The evolution equation of fluctuation including quantum gravity effects was first discussed in \cite{mc,starobinsky2}. But, in those days, the Riegert action and the running coupling effect were not taken into account.
} 
These equations connect the CFT phase with the Friedmann phase as a function of the running coupling constant. Within the linear approximation, it was shown that the amplitudes of Bardeen's gravitational potentials defined by $ds^2=a^2[-(1+2\Psi)d\eta^2+(1+2\Phi)d{\bf x}^2]$ are reduced small enough during the inflation.\footnote{ 
The behavior is different from the ordinary inflation driven by the cosmological constant in which the gravitational potential decays exponentially. 
} 
The result is depicted in fig.\ref{evolution of scalar mode}. The initial fluctuation is prepared by that of the conformal mode $\vphi$ with the condition $\Phi=\Psi$, while two scalar potentials mix during the evolution and at the transition point the dynamics drives the potentials to $\Phi=-\Psi$, which connect to the Friedmann universe. The amplitude of order of $10^{-5}$ is obtained numerically by choosing parameters in the factor (\ref{dynamical factor}) which may depend on strong coupling dynamics to be the values resulting in the proper number of e-foldings as shown in fig.\ref{evolution of scale factor}. This result agrees to the rough estimate using the ratio of two mass scales as discussed in Section 2.

Now, we discuss how the scalar spectrum changes during the inflation when non-linear effects are taken into account. It was known that the non-linear term in (\ref{expansion of vphi_NL}) tends to shift the scalar spectral index toward blue side from the Harrison-Zel'dovich-Peebles spectrum by the order of the initial amplitude $1/b_1$ \cite{amm97,hy}. It was shown that within the linear approximation the initial pattern of the spectrum holds for the comoving momentum less than $m$, but it becomes impossible to preserve the power-law behavior for the high momentum region beyond $m$ \cite{hhy}. Indeed, the non-linear effect acts to guarantee the power-law behavior of CFT beyond the Planck scale, and we can expect that the amplitude of scalar fluctuation will get smaller while preserving the power-law behavior during the inflation under the non-linear effect.

Although the non-linear distortion of the scalar spectrum is significant initially reflecting that the initial amplitude of $\vphi$ is not so small as to be neglected, we can expect from the observation above that a scale-invariant Harrison-Zel'dovich-Peebles spectrum is realized after the inflation as follows. The amplitude will reduce during the inflation preserving the power-law behavior, while the relationship between the linear and non-linear terms reflected in the order of $f_{\rm NL} \simeq 1$ will be fixed as a consequence of the diffeomorphism invariance (see also \cite{pc}). It implies that, as the scalar amplitude decreases to a small value of order $A_s ~(\simeq 10^{-10})$ at the transition time, the non-linear terms become negligible to order of square of the amplitude. Therefore, the scalar spectrum will reduce to the Harrison-Zel'dovich-Peebles spectrum $(n_s \to 1)$, which is an original prediction of the dimensionless field governed by the four-derivative dynamics of conformal gravity.

Let us discuss the dynamical effect which is originated from the existence of the correlation length $\xi_\Lam ~(\gg L_\P)$. This length scale implies the lack of correlation between two points separated beyond the correlation length $\xi_\Lam$ at the initial stage before the universe starts inflating. We incorporate this fact to the primordial spectrum through the spectral index which have a quantum correction of order of $t_r^2$ \cite{hy}. In order to take into account the non-perturbative effect we replace the coupling constant by the running coupling constant $\bar{t}_r(k)$. Thus, we obtain the almost Harrison-Zel'dovich-Peebles spectrum with the sharp fall-off at $k=\lam$, expressed as
\begin{equation}
     P_s(k)= A_s \left(\fr{k}{m} \right)^{v/\log(k^2/\lam^2)},
      \label{scalar spectrum at transition}
\end{equation}
where we leave the constant $v$ as a free parameter and the comoving dynamical scale is defined by
\begin{equation}
      \lam = a(\tau_i) \Lam_\QG 
      \label{comoving dynamical scale}
\end{equation}
similar to the case of the comoving Planck mass scale, and thus $H_\D/\Lam_\QG=m/\lam$. This scale indicates that there is no correlation for fluctuations of the spatial distance larger than $1/\lam$ today.

Since we use the comoving wave number, the physical scale of the fluctuation we consider here becomes much larger than the correlation length at the transition so that the pattern of the spectrum will be printed in the classical universe without modification by the strong coupling dynamics at the phase transition. Thus, the spectrum (\ref{scalar spectrum at transition}) gives the primordial scalar spectrum to compute the CMB angular power spectra.

\paragraph{Tensor spectrum}

The existence of the tensor mode implies a violation of the conformal invariance, which will be small initially because of the asymptotically free nature of this mode. The tensor spectrum at the Planck time is then given by the two-point correlation function of the transverse-traceless component $h^{\T\T}_{ij}$ of the traceless tensor field $h_{\mu\nu}$. It is also described by a dimensionless linear field with logarithmic correlation function. Unlike the scalar mode, the amplitude of the tensor mode $A_t$ is preserved during the inflation \cite{hhy} so that the spectrum at the transition point is given by
\begin{equation}
    P_t (k) = A_t \left( \fr{k}{m} \right)^{v/\log(k^2/\lam^2)},
             \label{tensor spectrum at transition}
\end{equation}
where $A_t$ is a small dimensionless constant. The spectrum corresponds to the tensor spectral index of $n_t =0$, apart from the damping factor originated from the dynamical scale, which is taken to be the same as that for the scalar spectrum. Thus, the tensor-to-scalar ratio $r=A_t/A_s$ will become significant at the transition point.

\section{CMB Angular Power Spectra}
\noindent

\begin{figure}[t]
\begin{center}
\includegraphics[width=11cm,clip]{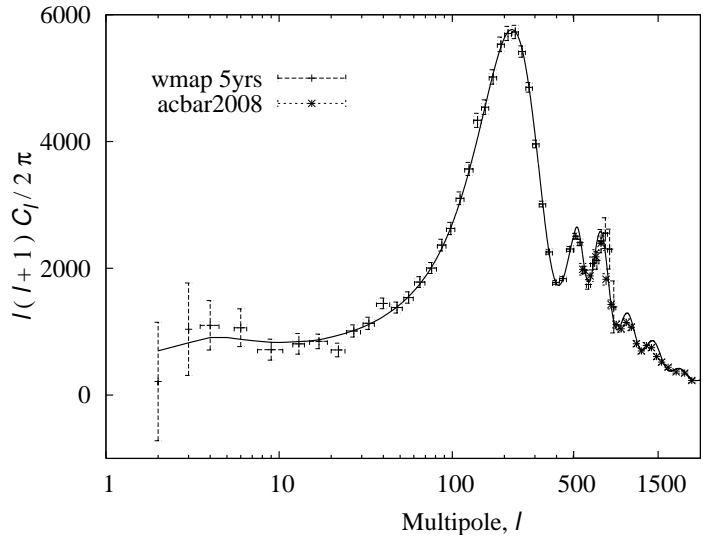}
\end{center}
\caption{\label{TT spectrum} {\small The TT spectrum with the WMAP5 and ACBAR data. The tensor-to-scalar ratio is taken to be $r=0.06$ and the parameters in the dynamical damping factor are chosen as $\lam=0.00026~(=m/60)$Mpc$^{-1}$ with $v=0.0002$. The optical depth is $\tau_e=0.08$, which is determined from the EE spectrum (not depicted), and the scalar amplitude is normalized appropriately. The other cosmological parameters are fixed to be $\Om_c=0.20$, $\Om_b=0.043$, $\Om_{\rm vac}=0.757$, $H_0=73$, $T_{\rm cmb}=2.726$, and $Y_{\rm He}=0.24$. The quality of fit is $\chi^2/{\rm dof} =1.10 ~(2 \leq l \leq 1000)$.}}
\end{figure}

We compute the CMB multipole components using the cmbfast code \cite{cmbfast} by taking the initial conditions to be the almost scale-invariant spectra (\ref{scalar spectrum at transition}) and (\ref{tensor spectrum at transition}) given at the transition point for the scalar and tensor modes, respectively.

The comoving dynamical scale is chosen to be $\lam = 0.00026$Mpc$^{-1}$ in order to explain the sharp damping at the $l=2$ multipole components. This value is consistent with the inflationary scenario given in fig.\ref{evolution scenario}, in which the scale factor grows up about $10^{59}$ from the Planck time to today, and thus we can expect from the definition (\ref{comoving dynamical scale}) that the comoving dynamical scale becomes this order by taking $\Lam_\QG \simeq 10^{17}$GeV as estimated in Section 2.

\begin{figure}[t]
\begin{center}
\includegraphics[width=11cm,clip]{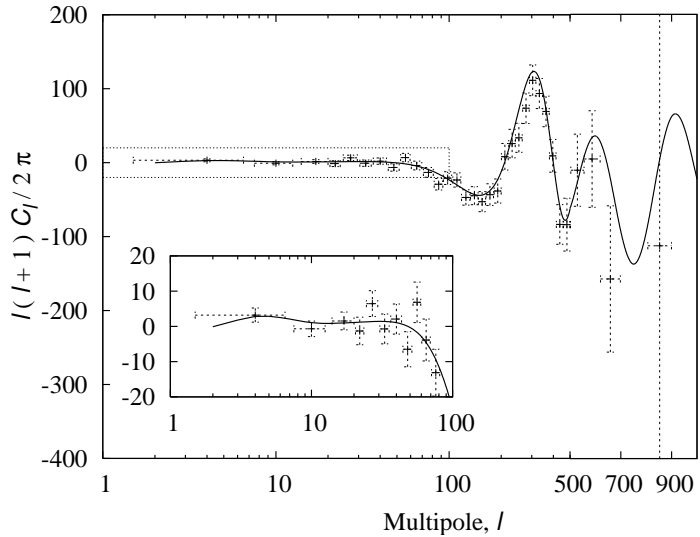}
\end{center}
\caption{\label{TE spectrum} {\small The TE spectrum with the WMAP5 data. The parameters are the same to those in fig.\ref{TT spectrum}. The quality of fit is $\chi^2/{\rm dof} =0.977 ~(2 \leq l \leq 1000)$.}}
\end{figure}

The tensor mode is necessary to compensate for lack of the amplitude in the multipole component lower than $100$. The tensor-to-scalar ratio is taken to be $r=0.06$ to fit with the WMAP5 data. The results for the temperature-temperature (TT) and temperature-polarization (TE) spectra are shown in figs.\ref{TT spectrum} and \ref{TE spectrum} together with the WMAP5 and ACBAR data, respectively.

We have assumed a unique decoupling time $\tau_\Lam$ for entire momentum range. However, if there is a time delay in the short scale at the space-time transition, the amplitude for high momentum region reduces further resulting red tilt of the spectrum. Data at high multipole components in future will tell us the detail of the space-time transition.

\section{Conclusion}
\noindent

We have studied the evolution of universe based on the renormalizable quantum gravity from the conformally invariant epoch before inflation to the transition point where the conformal invariance is completely broken. Introducing two characteristic mass scales, the number of e-foldings and the scalar amplitude at the transition point have been estimated as about $m_\pl/\Lam_\QG$ and $(\Lam_\QG/m_\pl)^2$, respectively. It suggests that the order of dynamical energy scale $\Lam_\QG$ locates at energy lower than $m_\pl$ by two orders.

We gave a special attention on the non-linear effects in the power-law scalar spectrum of CFT from the viewpoint of diffeomorphism invariance. The local non-Gaussianity parameter $f_{\rm NL}$ was evaluated to be the order of unity. Since the relationship between the linear and non-linear terms reflected in this order originates from the diffeomorphism invariance, it is expected to be preserved during the inflation. It suggests that as the amplitude decreases in the inflation era the scalar spectrum approaches to a scale-invariant one predicted from four-derivative dynamics of dimensionless gravitational fields. In this way, we obtained the almost Harrison-Zel'dovich-Peebles spectrum for the entire range aside from the large scale beyond the correlation length.

The sharp damping factor at the mass scale $\lam = 0.00026$Mpc$^{-1}$ in the comoving frame originated from the dynamical scale $\Lam_\QG \simeq 10^{17}$GeV, which reflects the initial state when there is no correlation larger than the correlation length $\xi_\Lam=1/\Lam_\QG$ before the space-time experiences inflation. Thus, the ratio $\Lam_\QG/\lam \simeq 10^{59}$ represents the full scale factor of the universe growing up from about the Planck time before the inflation to today. This order is consistent with the inflationary scenario of the universe assuming the existence of the space-time transition occurring at the energy scale of $\Lam_\QG$.

By making use of the spectrum at the transition point for the initial condition of the cmbfast code, we calculated the CMB angular power spectra. The initial spectrum for tensor mode was also given by the scale-invariant spectrum with the damping factor. We took the tensor-to-scalar ratio to be $r=0.06$.  Results are compared to the WMAP5 and ACBAR data with a significant quality of $\chi^2$.

We wish to thank Naoshi Sugiyama for helpful discussions on analysis of CMB multipoles.

\vspace{5mm}


\appendix

\vspace{5mm}
\begin{center}
   {\Large {\bf Appendix}}
\end{center}

\section{Evolution Equations of the Bardeen Potentials}
\noindent

The Bardeen potentials are given by $\Phi=\vphi+h/6$ and $\Psi=\vphi-h/2$ in the longitudinal gauge, where $h=h_{00}$ is the time-time component of the traceless tensor mode. The evolution equation is constructed from the effective action described by the combined system of the Riegert, the Weyl and the Einstein actions with the dynamical effects of the running coupling constant. In order to obtain the equations of motion only for these variables, the combinations of equations independent of the stress-tensor of matter fields are considered and the following coupled equations are obtained \cite{hhy}: 
\begin{eqnarray}
  && \fr{b_1}{8\pi^2} B_0(\tau) \biggl\{ 
         -2 \pd_\eta^4 \Phi                   -2 \pd_\eta \hphi \pd_\eta^3 \Phi
         + \left( -8 \pd_\eta^2 \hphi + \fr{10}{3} \lap3 \right) \pd_\eta^2 \Phi 
             \nonumber \\ 
  && \qquad\qquad
         + \left( -12 \pd_\eta^3 \hphi +\fr{10}{3} \pd_\eta \hphi \lap3 \right) \pd_\eta \Phi  
         + \left( \fr{16}{3} \pd_\eta^2 \hphi  - \fr{4}{3} \lap3 \right) \lap3 \Phi
             \nonumber \\ 
  && \qquad\qquad 
         + 2 \pd_\eta \hphi \pd_\eta^3 \Psi 
         + \left(8 \pd_\eta^2 \hphi + \fr{2}{3} \lap3 \right) \pd_\eta^2 \Psi           
         + \left( 12 \pd_\eta^3 \hphi - \fr{10}{3} \pd_\eta \hphi \lap3 \right) \pd_\eta \Psi
             \nonumber \\ 
  && \qquad\qquad
         + \left( - \fr{16}{3} \pd_\eta^2 \hphi  - \fr{2}{3} \lap3 \right) \lap3 \Psi
    \biggr\}
             \nonumber \\ 
  && 
    + \fr{m_\pl^2}{8\pi} e^{2\hphi} \Bigl\{ 
           6\pd_\eta^2 \Phi                 + 18 \pd_\eta \hphi \pd_\eta \Phi 
           - 4 \lap3 \Phi                   -6 \pd_\eta \hphi \pd_\eta \Psi 
             \nonumber \\ 
  && \qquad\qquad\qquad    
         + \left( 12 \pd_\eta^2 \hphi + 12 \pd_\eta \hphi \pd_\eta \hphi - 2\lap3 \right) \Psi 
           \Bigr\}  = 0 
         \label{linear scalar equation}
\end{eqnarray}
and
\begin{eqnarray}
   && \fr{2}{\bar{t}_r^2(\tau)} \left\{ 4 \pd_\eta^2 \Phi -\fr{4}{3} \lap3 \Phi 
                            -4 \pd_\eta^2 \Psi + \fr{4}{3} \lap3 \Psi \right\}
             \nonumber \\ 
   && +\fr{b_1}{8\pi^2} B_0 (\tau) \biggl\{
          \fr{4}{3} \pd_\eta^2 \Phi           + 4 \pd_\eta \hphi \pd_\eta \Phi
          + \left( \fr{28}{3} \pd_\eta^2 \hphi -\fr{8}{3} \pd_\eta \hphi \pd_\eta \hphi 
                    -\fr{8}{9} \lap3 \right) \Phi
             \nonumber \\ 
   && \qquad\qquad\qquad
         - \fr{4}{3} \pd_\eta \hphi \pd_\eta \Psi 
         + \left( -\fr{4}{3} \pd_\eta^2 \hphi + \fr{8}{3} \pd_\eta \hphi \pd_\eta \hphi 
                   - \fr{4}{9} \lap3  \right) \Psi
     \biggr\}
            \nonumber \\
  && + \fr{m_\pl^2}{8\pi} e^{2\hphi} \left\{
          - 2 \Phi - 2 \Psi  \right\}  =0,
           \label{linear scalar constraint}
\end{eqnarray}
where $\lap3 =\pd_i \pd^i$ and $\hphi$ is an inflationary solution of homogeneous equation of motion (\ref{homogeneous evolution equation}). The second equation is of a second order obtained by factoring out the operator $\lap3$.

The dynamical factors $B_0(\tau)$ in both equations and the inverse of $\bar{t}^2_r(\tau)$ in the second equation, which come from the Riegert and the Weyl actions respectively, show that the conformal gravity dynamics disappears at the transition, and thus the Einstein gravity dominates in the equation of motion. The second equation then plays an important role in connecting between the inflation and the Einstein phases, namely initially in the limit $\bar{t}_r \rightarrow 0$ the conformal mode dominates such that $\Phi=\Psi$, while at the transition point where the coupling diverges, the configuration with $\Phi = -\Psi$ should be realized.


\end{document}